\documentclass[conference,letter]{IEEEtran}
\IEEEoverridecommandlockouts

\usepackage{mathrsfs}
\usepackage{amsfonts}
\usepackage{array}       
\usepackage{graphicx}
\usepackage{multirow}   
\usepackage{epstopdf}
\usepackage{stmaryrd}
\usepackage{amsthm,amssymb,amsmath}
\usepackage{multirow,url}
\usepackage{color,cite}
\usepackage{slashbox}
\usepackage[lined,vlined,ruled,commentsnumbered]{algorithm2e}

\newtheorem{theorem}{Theorem}

\newtheorem{proposition}{Proposition}

\ifodd 1211

\newcommand{\revv}[1]{{#1}}
\newcommand{\comm}[1]{\textbf{\color{red} (Comment: #1) }}
\newcommand{\rev}[1]{#1}

\newcommand{\com}[1]{\textbf{\color{red} (Comment: #1) }}
\newcommand{\comg}[1]{\textbf{\color{green} (COMMENT: #1)}}
\newcommand{\response}[1]{\textbf{\color{green} (RESPONSE: #1)}}
\else
\newcommand{\revv}[1]{#1}
\newcommand{\comm}[1]{}
\newcommand{\rev}[1]{#1}

\newcommand{\com}[1]{}
\newcommand{\comg}[1]{}
\newcommand{\response}[1]{}
\fi

\def\dd{\mathrm{d}}
\def\eq{\triangleq}

\def\sumN{\sum_{n=1}^N}
\def\sumM{\sum_{m=1}^N}

\def\sumTa{\sum_{\t=1}^T}
\def\sumZ{\sum_{z=1}^Z}

\def\T{\mathcal{T}}
\def\N{\mathcal{N}}

\def\tseg{\beta}
\def\btseg{\boldsymbol{\tseg}}
\def\R{\mathcal{R}}
\def\Q{Q}

\def\dset{\boldsymbol{S}}
\def\ds{\boldsymbol{s}}

\def\du{u}
\def\dr{r}
\def\dz{z}
\def\dts{t^{\mathrm{s}}}
\def\dte{t^{\mathrm{e}}}
\def\dunk{\du_{n[k]}}
\def\drnk{\dr_{n[k]}}
\def\dznk{\dz_{n[k]}}
\def\dtsnk{\dts_{n[k]}}
\def\dtenk{\dte_{n[k]}}

\def\rset{\widehat{\boldsymbol{S}}}
\def\rs{\hat{\boldsymbol{s}}}

\def\ru{\hat{u}}
\def\rr{\hat{r}}
\def\rz{\hat{z}}
\def\rts{\hat{t}^{\mathrm{s}}}
\def\rte{\hat{t}^{\mathrm{e}}}
\def\runk{\ru_{n[k]}}
\def\rrnk{\rr_{n[k]}}

\def\rtenk{\rte_{n[k]}}
\def\rumk{\ru_{m[k]}}
\def\rrmk{\rr_{m[k]}}
\def\rzmk{\rz_{m[k]}}
\def\rtsmk{\rts_{m[k]}}
\def\rtemk{\rte_{m[k]}}

\def\buf{q}
\def\h{h}

\def\SW{W}
\def\Pay{P}
\def\Ut{U}
\def\Va{V} 
\def\va{\theta}
\def\Lossdeg{L^{\textsc{qdeg}}}	
\def\ld{\phi^{\textsc{qdeg}}}
\def\Lossrebuf{L^{\textsc{rebuf}}}	
\def\lr{\phi^{\textsc{rebuf}}}

\def\Cost{C}
\def\Costc{E^{\textsc{cell}}}
\def\Costw{E^{\textsc{wifi}}}

\def\cct{c^{\textsc{time}}}
\def\ccv{c^{\textsc{data}}}
\def\cwt{w^{\textsc{time}}}
\def\cwv{w^{\textsc{data}}}

\def\xSW{\widetilde{\SW}}
\def\xPay{\widetilde{\Pay}}

\def\xVa{\widetilde{\Va}} 
\def\xLossdeg{\widetilde{L}^{\textsc{qdeg}}}	
\def\xLossrebuf{\widetilde{L}^{\textsc{rebuf}}}	

\def\xCostc{\widetilde{E}^{\textsc{cell}}}
\def\xCostw{\widetilde{E}^{\textsc{wifi}}}

\def\SWo{\SW^{*}}
\def\SWx{\xSW^{*}}

\def\t{\tau}

\def\kvec{\boldsymbol{K}}		
\def\k{\kappa}
\def\x{x}		
\def\xd{\x^{\textsc{dl}}}
\def\y{y}		
\def\yr{\y^{\textsc{re}}}

\begin{document}

\title{Performance Bound Analysis for Crowdsourced Mobile Video Streaming \vspace{-2mm}
}


\author{Lin~Gao, Ming~Tang, Haitian~Pang, Jianwei~Huang, and Lifeng~Sun
\IEEEcompsocitemizethanks{
\IEEEcompsocthanksitem
 This work is supported by the General Research Fund (Project Number CUHK 14202814) established under the University Grant Committee of the Hong Kong Special Administrative Region, China, 
 and the National Natural Science Foundation of China (NSFC under No.~61472204).
\IEEEcompsocthanksitem
L.~Gao is with the School of Electronic and Information Engineering, Harbin Institute of Technology Shenzhen Graduate School, China, E-mail: gaolin@hitsz.edu.cn;
M.~Tang, J.~Huang, and L.~Gao are with the Network Communications and Economics Lab (NCEL),
Department of Information Engineering, The Chinese University of Hong Kong, HK,
E-mail: \{mtang, jwhuang, lgao\}@ie.cuhk.edu.hk;
H.~Pang and L.~Sun are with the Department of Computer Science and Technology, Tsinghua University, China, E-mail:
pht14@mails.tsinghua.edu.cn, sunlf@tsinghua.edu.cn.~~~~~~
}\vspace{-5mm}
}


\maketitle

\begin{abstract}
Adaptive bitrate (ABR) streaming enables video users to \emph{adapt} the playing bitrate to the real-time network conditions to achieve the desirable quality of experience (QoE).
In this work, we propose a novel   \emph{crowdsourced streaming framework} for multi-user ABR video streaming over wireless networks.
This framework enables the nearby mobile video users to crowdsource their radio links and resources for cooperative video streaming.
We focus on analyzing the social welfare performance \emph{bound} of the proposed crowdsourced streaming system.
Directly solving this bound is challenging due to the asynchronous operations of users.
To this end, we introduce a \emph{virtual} time-slotted system with the synchronized operations, and formulate the associated~social welfare optimization problem as a linear programming.
We   show that the optimal social welfare performance of the virtual system provides   effective upper-bound and lower-bound for the optimal performance (bound) of the original asynchronous system, hence characterizes the feasible performance \emph{region}  of the proposed crowdsourced streaming system.
The performance bounds derived in this work can serve as a benchmark for the future online algorithm design and incentive mechanism design.
\end{abstract}


%
\IEEEpeerreviewmaketitle

\addtolength{\abovedisplayskip}{-1mm}
\addtolength{\belowdisplayskip}{-1mm}


\vspace{-2mm}

\section{Introduction}

\subsection{Background and Motivations}

\emph{Adaptive BitRate (ABR)} streaming \cite{abr} is a widely-used technology for video streaming over large distributed HTTP networks such as Internet.
The key idea is to enable video players to \emph{adapt} the playing bitrate (which corresponds to the video quality such as resolution) to the real-time network conditions.
To achieve the flexible bitrate adaptation, a source video is first partitioned into a sequence of short multi-second \emph{segments}, each encoded at multiple pre-defined bitrates.
Then,
the bitrate adaptation of each video user can be achieved
by choosing different bitrates for different segments.
Clearly, with   proper bitrate adaptations, video users can achieve the desirable tradeoff between the quality of experience (QoE) and the streaming cost (e.g.,
energy consumption).

%
%
%
While most of the existing work on ABR streaming focused on the bitrate adaptation of a \emph{single user} \cite{b4,b1,b6,b2, b8,b5 ,b10,b7}, in this work we consider a more general scenario of \emph{multi-user} video streaming over wireless cellular networks.
Note that in a multi-user scenario, the QoE of each video user is affected not only by the stochastically changing
of his own network condition (such as wireless channel fading),
but also by the potential resource competition and interference of
other users \cite{comp1}.
Without proper coordination or cooperation among users, such  resource competition and multi-user interference may greatly degrade the network condition (e.g., leading to network congestion),
hence increase the streaming cost (due to, for example, the increased transmission power or repeated data retransmissions) and degrade the QoE of video users.
However, traditional single-user based bitrate adaptation methods in \cite{b4,b1,b6,b2, b8,b5 ,b10,b7} often fail to provide a desirable QoE for video users in the multi-user scenario, due to the lack of considerations of the potential cooperation among video users.

\begin{figure}[t]
  \centering
 \includegraphics[scale=0.65]{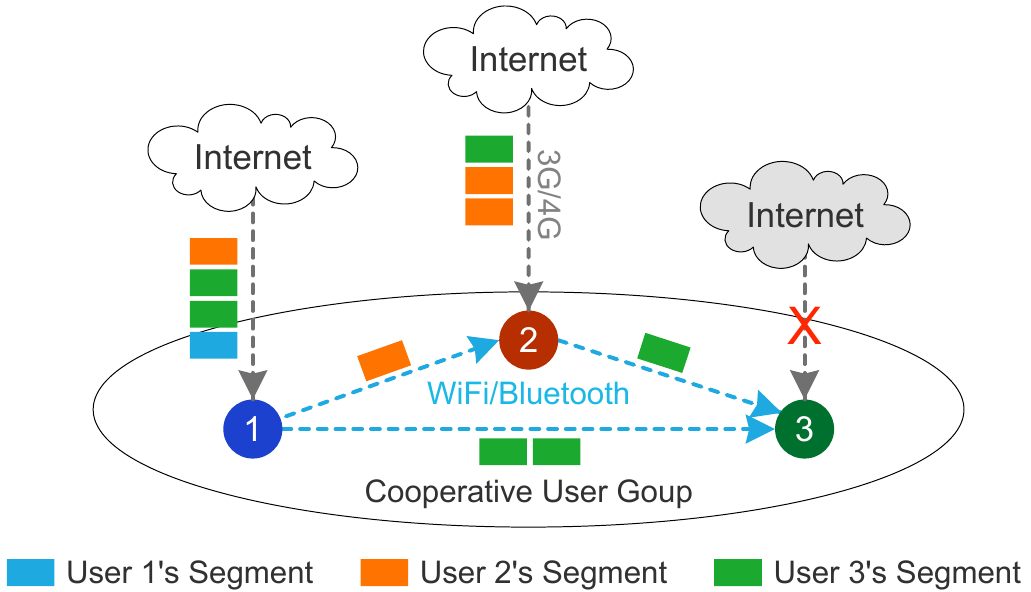}
  \vspace{-1mm}
  \caption{Crowdsourced Video Streaming Model.}
  \label{fig:model}
  \vspace{-3mm}
\end{figure}

In this work, we propose a novel user cooperation framework, called \emph{{crowdsourced (video) streaming}},
for multi-user video streaming over wireless cellular networks,
based on the \emph{user-provided networking} (UPN)  technology \cite{upn-lin,opengarden-lin,upn-infocom}.
The key idea is to enable nearby mobile users to form a cooperative group (via WiFi or Bluetooth) and crowdsource their cellular radio connections and resources for cooperative video streaming.\footnote{The idea of crowdsourcing has also been applied in other applications such as wireless community networking \cite{lin-wifi} and mobile crowdsensing \cite{lin-ps}.}
Namely, in a cooperative group, each user can download video segments for other users using his cellular link and resources and download his own video segments through other users' links and resources.
Figure \ref{fig:model} illustrates such a crowdsourced streaming model with a cooperative user group \{1, 2, 3\}, where user 1 downloads one segment for user 2 and two segments for user 3 (who has no available cellular link), and user 2   downloads one segment for user 3.

\subsection{Solution and Contributions}

Specifically, we focus on studying the users' cooperative streaming operations (including \emph{download scheduling} and \emph{bitrate adaptation})
and analyzing the theoretical social welfare performance \emph{bound} of such a crowdsourced streaming system.
Namely, for each video user, when and for whom he is going to download the video segments at which bitrates, for the purpose of maximizing the social welfare performance?

We first formulate the users' cooperative streaming operations in the crowdsourced system and the associated social welfare optimization problem.
The solution of such a problem can provide the theoretical social welfare performance bound of the crowdsourced system.
However, directly solving this bound is challenging due to the asynchronous operations of users as well as the mixed-integer nature of the problem.~~~~~~~~

To this end, we introduce a \emph{virtual} time-slotted system with  the synchronized operations, and formulate the associated  social welfare optimization problem as a linear programming (which can be solved efficiently with many standard methods).
We   show that with proper choices of time parameters, the optimal social welfare performance of the virtual time-slotted system provides effective upper-bound and lower-bound for the optimal performance (bound) of the original asynchronous system,
which leads to the feasible performance region of the proposed crowdsourced streaming system.
In summary, we list the key contributions of this work as follows.

\begin{itemize}
\item \emph{Novel Model:}
To our best knowledge, this is the first work that proposes a crowdsourced streaming framework for multi-user cooperative video streaming.
\revv{This framework enables mobile video users to crowdsource their radio connections and resources for cooperative video streaming,
hence can increase the users' QoE.}


\item \emph{Performance Bound Analysis:}
We analyze the theoretical social welfare performance bound of the proposed crowdsourced system comprehensively, overcoming the challenging issue of asynchronous operations by introducing a virtual time-slotted system.

\item \emph{Practical Insights:}
The performance bound analysis in this work is an important first-step towards the online algorithm design and incentive mechanism design for a crowdsourced streaming system, where the performance bounds in this work can serve as a benchmark.


%

\end{itemize}

The rest of the paper is organized as follows.
In Section II, we present the system model.
In Section III, we provide the problem formulation.
In Section IV, we propose the virtual time-slotted system and the performance bound analysis.
In Section V, we conclude this work and discuss the future work.


\section{System Model}

\vspace{-1mm}
\subsection{Network Model}


We consider a set $\N \eq \{1,2,...,N\}$ of mobile video users, and each user desires to watch a video (on his smartphone) via wireless cellular network.
Users move randomly in a certain area, and   nearby users can form a cooperative group (via WiFi) and crowdsource their radio connections and resources for cooperative video streaming.\footnote{We assume that some well-designed incentive mechanisms (e.g., Nash bargaining \cite{nash}) have been adopted such that all users are willing to participate in such a crowdsourced system to help others.} 
We refer to such a multi-user cooperative video streaming scheme as \emph{crowdsourced (video) streaming}.
Figure \ref{fig:model} illustrates such a crowdsourced streaming model with a cooperative user group \{1, 2, 3\}.

We consider the operation in a period of continuous time $\T \eq  [0,\ T]$, where $t = 0$ is the initial time and $t= T $ is the ending time.
Let $h_n(t) > 0$ denote the cellular link capacity of user $n$ at time $t$, and $e_{n,m}(t) \in \{0,1\}$ denote whether users $n$ and $m$ are close enough at time $t$ such that they can connect with each other via WiFi (hence help each other).
We refer to $\{( h_n(t), e_{n,m}(t)), \forall m,n\in\N,t\in\T \}$ as the \emph{network information}, which varies randomly over time.

\subsection{Video Streaming Model}




We  consider a typical ABR streaming
standard \cite{abr},
 where a single source video file is partitioned into multiple segments and delivered to a video user using HTTP.
The key features of this ABR streaming model are summarized below.

(i) \emph{Video Segmenting:}
A source video file is divided into a sequence of small segments, each containing a short interval of playback time (e.g., 2--10 seconds) of the source video, which is possibly several hours in term of the total duration.
A user downloads the video segment by segment.

(ii) \emph{Multi-Bitrate Encoding:}
Each segment is encoded at multiple bitrates, each corresponding to a specific quality such as resolution.
A user can select different bitrates for different segments according to real-time network conditions.

(iii)  \emph{Data Buffering:}
Each downloaded segment is first saved in a buffer (e.g., 20--40 seconds) at the user's device, and then fetched to the video player sequentially for playback.

We denote $\tseg_{n}>0$ as the segment length (in seconds) of user $n$'s video, $ \Q_n > 0$ as the maximum buffer size (in seconds) of user~$n$'s device, and $\R_n \eq \{R_n^1, ..., R_n^Z\}$ as the set of bitrates (in Mbps) available for user $n$, which depends on both the sever-side protocols and the user-side parameters such as device type and video player.

%
%
%
%
%
%
%
%


\section{Problem Formulation}\label{sec:formulation}

In this section, we characterize the users' cooperative video downloading operations in the crowdsourced streaming model, and formulate the associated optimization problem.

Specifically, with the ABR streaming, each source video is downloaded \emph{segment by segment}.
Namely, each user starts to download a new segment (with a specific bitrate) only when completing the existing segment downloading.
Hence, users operate in an \emph{asynchronous} manner, as they may complete  segment downloading at different times.
We refer to such an operation scheme as the \emph{segmented}  download operation.

\subsection{Downloading Sequence}
With the segmented operation, each user $n$'s downloading operation can be characterized by a sequence:
\begin{equation}
\dset_n \eq \left\{\ds_{n[1]},\ \ds_{n[2]},\ ... ,\ \ds_{n[k]},\ ... \right\},
\end{equation}
with each element $\ds_{n[k]} $ denoting the information of the $k$-th downloaded segment,   including
the segment owner $\du $,
bitrate level $\dz $,
bitrate $\dr = R_{\du}^{\dz}$,
download start time $\dts $,
and end time $\dte $.
Namely, we can write $\ds_{n[k]} $ as a tuple
\begin{equation*}
\ds_{n[k]} = \left(\du ,\ \dz ,\ \dr,\ [\dts , \dte]  \right).
\end{equation*}
To distinguish different segments, we will also write $\ds_{n[k]}$ as $\{\dunk,\dznk,\drnk,\dtsnk,\dtenk\}$ when needed.


Next we provide the   constraints for a \emph{feasible} downloading sequence $\dset_n $ of user $n$.

(i) \emph{Timing Constraint:}
As users download videos segment by segment, we have the following timing constraint:
\begin{equation*}
\mathrm{C.1:}\quad
 \dtenk  \leq \dts_{n[k+1]} ,\quad  \forall k=1,..., |\dset_n  |;
\end{equation*}
A strict inequality implies that user $n$ waits for some time before starting to download the next segment $ \ds_{n[k+1]} $, {e.g., when the buffers of all users are full (see Section \ref{sec:form-receive}).}


(ii) \emph{Capacity Constraint:} Each segment $\ds_{n[k]} =  (\du ,  \dz, \dr ,$ $  [\dts ,  \dte]   )$
consists of $\dr \cdot \tseg_{\du}    $ Mbits of video data, and is downloaded by user $n$ within time interval $\big[\dtsnk , \dtenk \big]$.
Hence, we have the following cellular link  capacity    constraint:
\begin{equation*}
\mathrm{C.2:}\quad
r \cdot  \tseg_{\du} \leq \int_{\dtsnk }^{\dtenk } \h_n(t) \dd t ,\quad  \forall k=1,...,|\dset_n|,
\end{equation*}
where $\h_n(t)$ is the real time cellular link capacity (in Mbps) of user $n$ at time $t$, and changes with time.

(iii) \emph{Encounter  Constraint:}
Each user can only download data for nearby ``encountered'' users.
Hence, a segment with $\ds_{n[k]} =  (\du , \dz, \dr ,  [\dts ,  \dte]   )$ with $n\neq \du$ is feasible only if users $n$ and $\du $ are encountered during the interval $\big[\dtsnk ,\ \dtenk \big]$, i.e.,
\begin{equation*}
\mathrm{C.3:}\quad
e_{n,\du}(t) =1, \ t \in \left[\dtsnk ,\ \dtenk \right], \quad \forall k=1,...,|\dset_n| .
\end{equation*}

\subsection{Receiving Sequence}  \label{sec:form-receive}

Given the feasible downloading sequences of all users, i.e., $\dset_n, \forall n\in\N$, we can derive the segment receiving sequence of each user $m  $ as follows:\footnote{{We do not consider the WiFi transmission time here, as the WiFi link capacity (typically tens to hundreds Mbps) is usually much larger than a video bitrate (typically low than than two Mbps).}}
\begin{equation}
\rset_m = \bigcup_{n\in\N,k\in\{1,\ldots,|\dset_n|\}:\dunk = m}  \left\{\ds_{n[k]}  \right\}
\end{equation}
We assume that a proper download scheduling has been adopted, such that there is no repeated segments within $\rset_m$, and all segments in $\rset_m$ are sorted according to the playback order.  We denote the $k$-th segment in the reordered $\rset_m $ by $\rs_{m[k]}$. Then, we can write the receiving sequence of user $m$~as:
\begin{equation}
\rset_m \eq \left\{\rs_{m[1]},\ \rs_{m[2]},\ ... ,\ \rs_{m[k]}, \ ... \right\},
\end{equation}
with each element $\rs_{m[k]} = \left(\ru ,\ \rz,\ \rr ,\ [\rts , \rte]  \right) $ denoting the information of the $k$-th segment played by user $m$.
Similarly, we will write $\rs_{m[k]}$ as $\{\rumk,\rzmk,\rrmk,\rtsmk,\rtemk\}$ when needed.
 It is easy to see that  $\rumk = m$  for all $\rs_{m[k]} \in \rset_m $.
To facilitate the later analysis, we further assume that $\rtemk \leq \rte_{m[k+1]}$, $\forall k=1,...,| \rset_m |$, that is, user $m$ receives the segments in $\rset_m$ sequentially.\footnote{Note that this can always be achieved by a proper schedule of downloading sequences with the full network information. For example, if $\rtemk > \rte_{m[k+1]} $, i.e., the $k+1$-th segment is received before the $k$-th segment, we can simply change their downloading orders.}


As mentioned previously, each received segment
is first stored in a buffer at the user's device, and then fetched to the video player sequentially for playback.
Let $\buf_{m[k]}$ denote the buffer level (in seconds) of user $m$ \emph{when receiving the $k$-th segment}, i.e., at the time $\rtemk$.
Then, we have the following \emph{\textbf{buffer update rule}} for user $m$:
\begin{equation}\label{eq:buffer-rule}
\buf_{m[k]} = \left[\buf_{m[k-1]} - \left(\rtemk - \rte_{m[k-1]} \right)\right]^+ + \tseg_m,
\end{equation}
where $[x]^+ = \max\{0,x\}$.
Here $\rtemk - \rte_{m[k-1]}$ is the time interval between   receiving of  $\rs_m[k-1] $ and $\rs_{m[k]} $, during which a period $\rtemk - \rte_{m[k-1]}$ of video is played back and removed from the buffer;
 $\tseg_m$ is the segment length (playback time) of the newly received segment $\rs_{m[k]}$.

Since each user $m$'s buffer size is limited at $Q_m$ (seconds), we have the following \emph{buffer constraint}:
\begin{equation*}
\mathrm{C.4:}\quad
0 \leq \buf_{m[k]} \leq  Q_m,\quad \forall k=1,...,| \rset_m |.
\end{equation*}

\vspace{-2mm}

\subsection{User Payoff}


The payoff   of a video user mainly consists of two parts: a \emph{utility} function capturing the user QoE for video service, \revv{and a \emph{cost} function capturing the user's energy consumption for both data downloading and data local exchanging.}

\emph{1) \textbf{Quality-of-Experience (QoE):}}
Users often desire  for a higher video quality without  frequent quality changes and freezes during playback.
Hence, a user's QoE mainly depends on the video quality,  quality fluctuation, and  rebuffering.

(i) \emph{Video Quality:}
A higher video  quality (bitrate) brings a higher QoE for users.
Let $g_n(r)$ denote the value that user $n$ achieves from  bitrate $r$ during one unit of playback time.\footnote{Precisely speaking, this value is a function of quality. Nevertheless, under the assumption that there is a distinct and monotonic relationship between bitrate and quality, we write it as function of bitrate  for notational convenience.}
Then, the total value that user $n$ achieves from all received segments $\rset_n$ (each with a playback time of $\tseg_{\runk} = \tseg_n$) is:
\begin{equation} \label{eq:value}
\Va_n (\rset_n) \eq \sum_{k=1}^{|\rset_n|} g_n \left( \rrnk \right) \cdot \tseg_{n}.
\end{equation}
Obviously, $g_n(\cdot)$ is an increasing function  (as video quality monotonically increases with bitrate).
As an example, we can adopt the following value function \cite{comp3}: $g_n(r) = \log(1 + \va_n \cdot r)$, where $\va_n >0 $ is a user-specific evaluation factor capturing user $n$'s desire for a high quality video service.

(ii) \emph{Quality Fluctuation:}
The change of quality (bitrate) during playback decreases the users' QoE, especially when the quality is degraded.
In this work, we assume that there is a value loss proportional to the bitrate decrease once the quality is degraded,
while there is  no value loss when the quality is upgraded \cite{comp3}.
Let $\ld_n > 0 $ denote the value loss of user $n$ for one unit (in Mbps) of bitrate decrease. Then, the total value loss of user $n$ induced by quality degradation is
 \begin{equation}\label{eq:loss-qd}
\Lossdeg_n (\rset_n) \eq \sum_{k=2}^{| \rset_n |} \ld_n \cdot \left[  \rr_{n[k-1]} -  \rrnk \right]^+,
\end{equation}
where $[x]^+ = \max\{0,x\}$. Here $\rr_{n[k-1]} >  \rrnk $ indicates that a quality degradation occurs between $\rs_n[k-1] $ and $\rs_{n[k]} $, with a bitrate decrease
 of  $\rr_{n[k-1]} -  \rrnk $.

(iii) \emph{Rebuffering:}
If a video buffer is exhausted before receiving a new segment, the video player   has to freeze the playback and rebuffer the video for a certain time. Such a freeze during playback is called \emph{rebuffering}.
The rebuffering (freeze) during playback greatly affects the users' QoE.
By the buffer update rule \eqref{eq:buffer-rule}, a rebuffering occurs when
\begin{equation*}
\buf_{n[k-1]} <  \rtenk - \rte_{n[k-1]},
\end{equation*}
with a detailed rebuffering time $\rtenk - \rte_{n[k-1]} - \buf_{n[k-1]}$.
Let $\lr_n >0 $ denote the value loss of user $n$ for one unit  of rebuffering time. Then, the total value loss of user $n$ induced by video rebuffering~is
 \begin{equation}\label{eq:loss-rebuf}
\Lossrebuf_n (\rset_n) \eq \sum_{k=2}^{| \rset_n |} \lr_n \cdot \left[ \rtenk -  \rte_{n[k-1]} - \buf_{n[k-1]} \right]^+.
\end{equation}

Based on the above, we can define the \emph{utility} of each user $n$ under a receiving sequence $\rset_n$   as follows:
\begin{equation}\label{eq:utility}
\Ut_n(\rset_n)\eq \Va_n (\rset_n)- \Lossdeg_n (\rset_n)- \Lossrebuf_n (\rset_n).
\end{equation}

\emph{2) \textbf{Energy Cost:}}
Users incur some energy cost in video streaming.
\rev{Such energy cost mainly includes the energy consumption for data downloading on cellular links, and energy consumption for data exchange over local WiFi links.}

(i) \emph{Energy Consumption for Video Downloading (via Celluar and Internet):}
When downloading data via the cellular link (and Internet), users' energy consumption depends on both the downloading time and the downloaded data volume \cite{energy}.
Let $\cct_n \geq 0$ denote the time-related energy consumption factor of user $n$ (i.e., for each unit  of downloading time), and $\ccv_n \geq 0$ denote the volume-related energy consumption factor of user $n$ (i.e., for each unit of downloaded data).
Then, the energy consumption of user $n$ for downloading video contents via cellular links and Internet is \cite{energy}:
\begin{equation*}\label{eq:cost-c}
\Costc_n (\dset_n) \eq \sum_{k=1}^{|\dset_n|}  \left(\cct_n \cdot (\dtenk - \dtsnk)
+ \ccv_n \cdot  \drnk \cdot \tseg_{\dunk}\right).
\end{equation*}

(ii) \emph{Energy Consumption for Video Exchanging (via WiFi):}
When downloading a segment for others, the user needs to transmit the data to the segment owner via the local WiFi link,
the energy consumption of which also depends on the transmitting time and the transmitted data volume \cite{energy}.
Let $\cwt_n \geq 0$ and $\cwv_n \geq 0$ denote the time-related and volume-related energy consumption factors of user $n$ on the WiFi link, respectively.
The   energy consumption of user $n$  for video exchanging on WiFi link is \cite{energy}:
\begin{equation*}\label{eq:cost-w}
\Costw_n (\dset_n) \eq \sum_{k=1}^{|\dset_n|}  \left(\cwt_n \cdot 0
+ \cwv_n \cdot \drnk \cdot \tseg_{\dunk}\right) \cdot \textbf{1}( \dunk\neq n ),
\end{equation*}
where the indicator function $\textbf{1}( \dunk\neq n ) =1 $ if $\dunk\neq n$ (i.e., the segment $\ds_{n[k]}$ is downloaded for others), and $0 $ otherwise.
Here we assume that the WiFi transmission time of a single segment is small and hence negligible.

Based on the above, we can derive the total \emph{energy consumption} of each user $n$ under a downloading sequence $\dset_n$ and receiving  sequence $\rset_n$ as follows:
\begin{equation}\label{eq:cost}
\Cost_n (\dset_n,\rset_n) \eq \Costc_n (\dset_n) + \Costw_n (\dset_n) .
\end{equation}

\emph{3) \textbf{Payoff:}}
The {payoff} of user $n$, denoted by $\Pay_n $, is defined as the difference between utility (capturing the QoE of users) and cost (capturing the energy consumption), i.e.,
\begin{equation}\label{eq:payoff}
\begin{aligned}
\Pay_n ( \dset_n , \rset_n )   \eq & \Ut_n (\rset_n)  - \Cost_n (\dset_n, \rset_n)
\end{aligned}
\end{equation}

The \emph{social welfare} is the aggregate payoff of all users, i.e.,
\begin{equation}\label{eq:sw}
\SW (\dset_1, ..., \dset_N) \eq \sumN \Pay_n (\dset_n, \rset_n),
\end{equation}
where $\rset_n $ can be derived from $\dset_n, \forall n\in\N$.

%

\subsection{Problem Formulation}


We consider an ideal scenario with \emph{complete} network information in this work, and formulate the following \emph{offline} social welfare maximization problem:\footnote{Note that without complete (future) network information, we cannot formulate this offline social welfare maximization problem.
In this case, we need to design online   algorithms, where the downloading operation of each user is performed in an online and distributed manner.}
\begin{equation}\label{eq:swm}
\begin{aligned}
\max_{\{\dset_n, n\in\N\}} ~~& \SW (\dset_1, ..., \dset_N),
\\
\mbox{s.t.} ~~& \mathrm{C.1\sim C.4}.
\end{aligned}
\end{equation}
The solution of \eqref{eq:swm}, denoted by $\SWo$, provides the theoretical performance bound (in term of social welfare) of the proposed crowdsourced system.
However, directly solving  \eqref{eq:swm} is very challenging due to the following reasons.
First, users operate in an asynchronous manner.
Namely, users may start to download new segments at different time.  
Second,  \eqref{eq:swm} involves both discrete variables  (e.g., $\du$ and $\dz$) and  continuous  variables (e.g., $\dts$ and $\dte$), hence is a complicated mixed-integral  optimization problem.
Third, \eqref{eq:swm} involves the   integral operation ($\mathrm{C.2} $), which makes it even more challenging  to solve.
Hence, in the next section, we will focus on finding upper-bound and lower-bound for this performance bound $\SWo$.




\section{Performance Bound Analysis}
\label{sec:main1}



In this section, we propose a virtual \emph{time-slotted download operation} scheme, under which the problem can be formulated as an linear programming, hence can be solved by many classic methods.
We will show that the solution of \eqref{eq:swm} under the segmented operation scheme (i.e., the theoretical performance bound of the proposed crowdsourced system) is bounded by the solutions under this virtual time-slotted system.
\textbf{It is important to note that this time-slotted operation scheme is only used for characterizing the theoretical performance bound, but not for the practical implementation.}

\vspace{-1mm}

\subsection{Time-Slotted Download Operation}


To model the time-slotted operation scheme, we divide the whole time period $[0, T]$ into multiple time slots, each with the same   length (e.g., 10 seconds).
For convenience, we normalize the length of each   slot to be one.
Hence, there is  a   set of $T$ time slots, denoted by $\T = \{1,2,...,T\}$, with the $\t$-th slot corresponding to   time interval $[\t-1,\ \t]$.

Under the time-slotted operation scheme, each video is downloaded \emph{slot by slot} in a {synchronized} manner, rather than segment by segment under the segmented operation.
Thus, in this case,  we can focus on the segments  that each user downloads in each time slot, instead of the segment downloading sequence.
Moreover, to guarantee the synchronous operation, we require that each segment must be completely downloaded within one time slot.
Namely, users cannot download a segment across multiple time slots.

\begin{figure}[t] 
  \centering
  \includegraphics[height=0.75in]{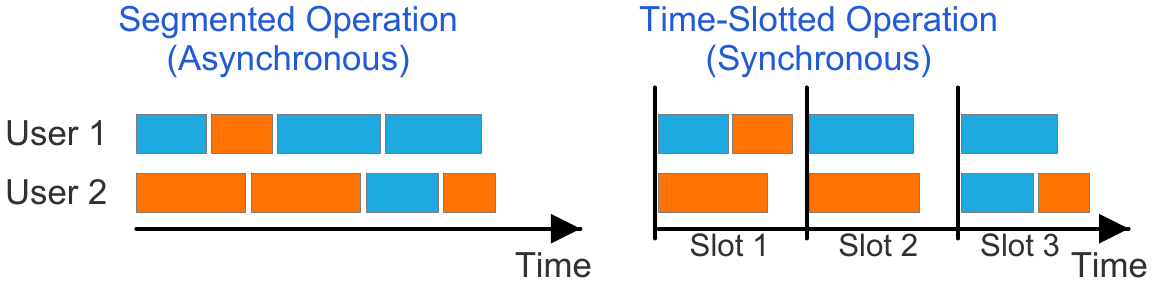}
\vspace{-2mm}
  \caption{Segmented vs Time-Slotted Operation.}
  \label{fig:operation}
\vspace{-4mm}
\end{figure}

For clarity, we illustrate the difference between the segmented operation   and the time-slotted operation in Figure \ref{fig:operation}. Blue blocks denote the user 1's data and orange blocks denote the user 2's data.
Under the segmented operation scheme (left), users start to download data at different times, while under the time-slotted
operation scheme (right), users are synchronized, and download data at the beginning of each time slot.

\emph{1) \textbf{Downloading Vector:}}
With the time-slotted   operation,
the downloading operation of each user $n$ can be characterized by a downloading vector:
\begin{equation}
\kvec_n \eq \left\{  \k_{n,m}^z(\t), \ \ \forall  \t \in\T,  m\in\N, z\in\{1,...,Z\}  \right\},
\end{equation}
where each element $\k_{n,m}^z(\t)  $ is a non-negative integer,  denoting the total number of segments with a bitrate level $z$ that user $n$ downloads for user $m$ in time slot $\t$.


Given the downloading vector $\kvec_n$,
we can derive the total amount of data that user $n$ downloads  in each time slot $\t$:
\begin{equation}
\xd_{n }(\t ) = \sumM  \x_{n,m}(\t) = \sumM    \sumZ \k_{n,m}^z(\t)  \cdot \tseg_m  \cdot  R_m^z ,
\end{equation}
where $\x_{n,m}(\t) \eq \sumZ \k_{n,m}^z(\t) \cdot \tseg_m \cdot  R_m^z $ is the amount of data   for user $m$ in slot $t$.
Then, we can define the link capacity constraint and encounter constraint for a feasible   $\kvec_n$:
\begin{equation*}
\begin{aligned}
\mathrm{\widetilde{C}.2:}\quad &
\xd_{n }(\t ) \leq  H_n({\t}),
\\
\mathrm{\widetilde{C}.3:}\quad &
e_{n,m}(t) = 1, t\in[\t-1, \t], \mbox{ if }	\x_{n,m}(\t) > 0,
\end{aligned}
\end{equation*}
where $H_n({\t}) =  \int_{\t-1 }^{\t } \h_n(t) \dd t$ is the aggregate  cellular link capacity (in Mbps) of user $n$ in time slot $\t$.
Note that with the time-slotted operation,  we do not need to consider the timing constraint (C.1), as the    operation is already     slot by slot.

 \emph{2) \textbf{Receiving Vector:}}
Given the feasible downloading vector of all users, i.e.,
$\kvec_n,\forall n\in N$, we can derive the total   playback time that user $m $ receives in each time slot $\t$:
\begin{equation}
\yr_m(\t)  =  \sumN \y_{n,m}(\t) = \sumN    \sumZ \k_{n,m}^z(\t) \cdot \tseg_m  ,
\end{equation}
where $\y_{n,m}(\t) \eq \sumZ \k_{n,m}^z(\t) \cdot \tseg_m$ is the total playback time that user $m$ receives from user  $n$ in slot $\t$.

Let $\buf_m(\t)$ denote the buffer level (in seconds) of user $m$ \emph{at the end of time slot $\t$}.
Then, we have the following \emph{\textbf{buffer update rule}} for user $m$:\footnote{Here one time unit of video is played back during time slot $\t$, and $\yr_m(\t)$ is the playback time of the newly received segments in slot $\t$.}
\begin{equation}\label{eq:buffer-rule-slot}
\buf_m(\t)  = \left[\buf_m(\t-1) - 1 \right]^+ + \yr_m(\t). 
\end{equation}

Similarly, we have the following buffer constraint:
\begin{equation*}
\mathrm{\widetilde{C}.4:}\quad
0 \leq \buf_m(\t) \leq  Q_m,\quad \forall \t=1,...,T.
\end{equation*}

 \emph{3) \textbf{User Payoff:}}
Now we define the user payoff and social welfare under the time-slotted operation.

(i) \emph{Video Quality:}
Similar as \eqref{eq:value}, the   value that user $n$ achieves from all received segments is:
\begin{equation} \label{eq:value-slot}
\xVa_n  \eq  \sumTa \sumM \sumZ \k_{m,n}^z(\t)  \cdot \tseg_n  \cdot g_n(R_n^z) .
\end{equation}

(ii) \emph{Quality Fluctuation:}
Without loss of generality, we assume that all the received segments of each user $n$ in each time slot $\t$ are sorted in the ascending order of bitrate.
Hence, quality degradation only occurs between two successive time slots, while never occurs within a time slot.
Let ${r}_n^{\textsc{h}}(\t)$ and $r_n^{\textsc{l}}(\t)$ denote the highest bitrate and lowest bitrate that user $n$ receives in slot $\t$.
 Then, similar as \eqref{eq:loss-qd}, the   value loss of user $n$ induced by quality degradation is
 \begin{equation}\label{eq:loss-qd-slot}
\xLossdeg_n \eq \sum_{\t=2}^{T} \ld_n \cdot \left[ r_n^{\textsc{h}}(\t-1) -  r_n^{\textsc{l}}(\t) \right]^+,
\end{equation}

(iii) \emph{Rebuffering:}
By the buffer update rule in \eqref{eq:buffer-rule-slot}, a rebuffering occurs in time slot $\t$ when
$$
\buf_m(\t-1) < 1 ,
$$
with a rebuffering time $1 - \buf_m(\t-1)$.
Then, similar as \eqref{eq:loss-rebuf}, the value loss of user $n$ induced by rebuffering~is
 \begin{equation}\label{eq:loss-rebuf-slot}
\xLossrebuf_n \eq \sum_{\t=2}^{T} \lr_n \cdot \left[ 1  -\buf_m(\t-1) \right]^+.
\end{equation}

(iv) \emph{Energy Consumption for Video Downloading (via Cellular and Interent):}
The energy consumption  of user $n$ for video downloading  on cellular link (and Internet)~is
\begin{equation}\label{eq:cost-c-slot}
\xCostc_n \eq \sumTa  \left(\cct_n \cdot \frac{\xd_{n }(\t )}{H_n({\t})}
+ \ccv_n \cdot  \xd_{n }(\t )  \right),
\end{equation}
where $\frac{\xd_{n }(\t )}{H_n({\t})}$ is the actual downloading time in time slot $\t$.

(v) \emph{Energy Consumption for Video Exchanging (via WiFi):}
The energy consumption of user $n$ for video exchanging on the local WiFi link is
\begin{equation}\label{eq:cost-w-slot}
\xCostw_n \eq \sumTa  \sum_{m=1, m\neq n}^N
\left(\cwt_n \cdot 0
+ \cwv_n \cdot \x_{n,m}(\t)   \right).
\end{equation}


Based on the above,  the payoff of each user $n$  is
\begin{equation}\label{eq:payoff-slot}
 \xPay_n (\kvec_1, ..., \kvec_N) \eq  \xVa_n - \xLossdeg_n - \xLossrebuf_n
- \xCostc_n - \xCostw_n.
 \end{equation}


 \emph{4) \textbf{Problem Formulation under Time-Slotted Operation:}}
 Now we can define the social welfare maximization problem under the time-slotted download operation as follows:
\begin{equation}\label{eq:swm-slot}
\begin{aligned}
\max_{\{\kvec_n, n\in\N\}} ~~ & \xSW   \eq \sumN \xPay_n(\kvec_1, ..., \kvec_N) ,
\\
\mbox{s.t.} ~~& \mathrm{\widetilde{C}.2\sim \widetilde{C}.4}.
\end{aligned}
\end{equation}
Similar to \eqref{eq:swm}, this is an \emph{offline} optimization problem and requires the complete network information.
Moreover, \eqref{eq:swm-slot} is an integer programming,
and can be solved by many classic methods.
Hence, we skip the detailed derivations.
For notation convenience, we denote the solution of \eqref{eq:swm-slot} by $\SWx$.

%

\subsection{Performance Bound}

Now we characterize the theoretical performance  bound $\SWo$ under the segmented operation, by using the solution $\SWx$ of \eqref{eq:swm-slot} under the virtual time-slotted operation.~~~~~~~

For convenience, we denote $\btseg \eq (\tseg_1, ..., \tseg_N)$ as the   vector consisting of all users' segment lengths, and denote  $\SWo_{(\btseg)}$ and  $\SWx_{(\btseg)}$
 as the    solutions of \eqref{eq:swm} and \eqref{eq:swm-slot} under  $\btseg$, respectively.
 {We refer to a   vector $\btseg$ as an \emph{integer multiple} of another vector $\btseg'$, if each element $\tseg_n$ in $\btseg$ is an integer multiple of the corresponding element $\tseg_n'$ in $\btseg'$}.
 For example, $\btseg = (1,...,N)$ is an
 integer multiple of $\btseg' = (0.5,...,N/2)$.

\begin{proposition}
If $\btseg$ is an  {integer multiple} of   $\btseg'$, then
$$
\SWo_{(\btseg)}  \leq  \SWo_{(\btseg')}, \mbox{~~~and~~~}
\SWx_{(\btseg)}  \leq  \SWx_{(\btseg')}.
$$
\end{proposition}

This proposition can be proved by showing that in both segmented and time-slotted operation schemes, any downloading operation under $ \btseg $ can be equivalently achieved under $\btseg'$.

\begin{proposition}
If $\btseg \rightarrow \mathbf{0}$ (i.e., $\tseg_n \rightarrow 0, \forall n\in\N$), then
$$
\SWo_{(\btseg)} = \SWx_{(\btseg)}.
$$
\end{proposition}

This proposition can be proved by showing that with infinitely small segment lengths $\btseg \rightarrow \mathbf{0}$, any downloading operation under the time-slotted operation can be equivalently achieved under the segmented operation, and vise versa.

\begin{proposition}
If $\btseg \succeq \mathbf{0}$ is a finite vector (i.e., each element $\tseg_n \geq 0$ is a finite number), then
$$
\SWo_{(\btseg)} \geq  \SWx_{(\btseg)}.
$$
\end{proposition}

This proposition can be proved by showing that with   finite segment lengths $\btseg \succeq \mathbf{0}$, any downloading operation under the time-slotted operation   can be equivalently achieved under the segmented operation, but \emph{not}   vise versa.

Based on the above, we have the following theorem.

\begin{theorem}
Given a segment length vector $\btseg $, the theoretical performance bound $\SWo_{(\btseg)}$ is bounded by:
$$
\SWx_{(\btseg)} \leq   \SWo_{(\btseg)} \leq  \SWx_{(\btseg' \rightarrow \mathbf{0})}.
$$
\end{theorem}

Intuitively, this theorem states that with any  $\btseg $, the theoretical performance bound $\SWo_{(\btseg)}$ of our proposed crowdsourced system is
(a) lower-bounded by $\SWx_{(\btseg)}$ (i.e., the optimal performance of the virtual time-slotted system with the same segment length vector $\btseg $), and   (b) upper-bounded by $\SWx_{(\btseg' \rightarrow \mathbf{0})}$ (i.e., the optimal performance of the virtual time-slotted system with infinitely small segment lengths $\btseg' \rightarrow \mathbf{0} $).
Therefore, the performance of the virtual time-slotted system under different $\btseg $ characterizes the theoretical performance region  of our proposed crowdsourced system.




\section{Conclusion}

In this work, we proposed a crowdsourced streaming framework for multi-user cooperative video streaming over mobile wireless networks, and analyzed the theoretical performance bound of the proposed crowdsourced streaming system.
There are two important directions for the future extension of this work.
First, it is important to study the online scheduling algorithms for the practical implementation of the proposed crowdsourced streaming system in the scenario without complete future and global network information.
Our performance bound analysis in this work can serve as a benchmark, and hence is an important first step towards the future online algorithm design for the crowdsourced streaming system.  
Second, incentive is a very important issue for a crowdsourced system, and is necessary for motivating video users participating the crowdsourced system to help others. 
Hence, it is also important to study the incentive issue in such a crowdsourced system. 
{More specifically, in the complete information scenario, this can be achieved by a Nash bargaining between the receiver and the downloader (in each segment downloading), with which each of them can achieve a welfare no worse than that in the non-cooperative system. 
In the incomplete information scenario, an incentive compatible mechanism (e.g., auction) is necessary to elicit the private information of users first, and then divide the generated social welfare properly among the receiver and the downloader.}


\end{document}